\documentclass[conference,compsoc]{IEEEtran}
\usepackage{url}
\usepackage{cite}
\usepackage{color}
\begin{document}
%
\title{Omnizart: A General Toolbox for Automatic Music Transcription}

\author{Yu-Te Wu, Yin-Jyun Luo, Tsung-Ping Chen, I-Chieh Wei, Jui-Yang Hsu, Yi-Chin Chuang, Li Su\\
Music and Culture Technology Lab, Institute of Information Science, Academia Sinica, Taipei, Taiwan
}


%


\maketitle



%

\section{Summary}
We present and release Omnizart, a new Python library that provides a streamlined solution to automatic music transcription (AMT). Omnizart encompasses modules that construct the life-cycle of deep learning-based AMT, and is designed for ease of use with a compact command-line interface. To the best of our knowledge, Omnizart is the first transcription toolkit which offers models covering a wide class of instruments ranging from solo, instrument ensembles, percussion instruments to vocal, as well as models for chord recognition and beat/downbeat tracking, two music information retrieval (MIR) tasks highly related to AMT. In summary, Omnizart incorporates:
\begin{itemize}
    \item Pre-trained models for frame-level and note-level transcription of multiple pitched instruments, vocal melody, and drum events;
    \item Pre-trained models of chord recognition and beat/downbeat tracking;
    \item Main functionalities in the life-cycle of AMT research, covering from dataset downloading, feature pre-processing, model training, to sonification of the transcription result. 
\end{itemize}

Omnizart is based on Tensorflow \cite{abadi2016tensorflow}. The code, command line interface, documentation, as well as demo examples can all be found in the project website:

\url{https://github.com/Music-and-Culture-Technology-Lab/omnizart} 

\section{Background}

AMT has been the core challenge in MIR because of the multifaceted nature of musical signals. Typically, streams of musical notes performed with various instruments overlap with each other and then create diverse levels of information abstraction in music; such a characteristic complicates the task to identify the attributes (e.g., instrument family, instrument class, pitch and pitch contour, duration, onset time, chord or non-chord tone, beat, etc.) of each note and each passage in the music. Previous studies mostly focus only on transcribing a restricted set of the attributes (e.g., piano solo transcription and other types of instrument-agnostic transcription), while a general solution applicable for all classes of instruments is rarely seen.

While the majority of the previous solution focuses on single-instrument transcription, Omnizart collects several SOTA models on the tasks of multiple pitched instrument transcription, drum transcription, vocal transcription, chord recognition and beat tracking, and serves as a unified toolkit that achieves transcription of multiple tracks and data modalities.
The pre-trained models in Omnizart closely reproduce the performances reported in the original papers, and can be fine-tuned with separate datasets for benchmarking purposes. In addition to music transcription, the transcribed outputs can also serve as the auxiliary information for other MIR tasks, e.g., source separation and music generation, especially in scenarios lacking supervisory resources. The release of Omnizart is expected to accelerate the advance of AMT research and contribute to the MIR community.

\section{Implementation details}


\subsubsection{Piano solo transcription} The piano solo transcription model in Omnizart reproduces the implementation of \cite{wu2020multi}. The model is a U-net structure that outputs a time-pitch representation with time resolution of 20ms and pitch resolution of 25 cents (1/4 semitone). The output time-pitch representation contains three 2-D channels, which are the pitch activation (i.e. piano roll) channel, the onset channel, and the offset channel. The MIDI transcription results are obtained with these output channels. The encoder and decoder parts of the U-net are constructed with stacked convolutional layers and residual layers following the implementation of DeepLabV3+ \cite{Chen2018DeepLabV3+}, and the bottleneck layer is a self-attention block which refers to the implementation of the Image Transformer \cite{parmar2018image}. The input of the model is a multi-channel data representation containing spectrogram, generalized cepstrum (GC) \cite{su2015combining} and the generalized cepstrum of the spectrogram (GCoS) \cite{wu2018automatic}. For better generalizability, the model is trained on the MAESTRO dataset \cite{hawthorne2018enabling}, an external dataset containing 1,184 real piano performance recordings with a total length of 172.3 hours. The number of parameters of the model is 7.92M. The model achieves 72.50\% of frame-level F1-score and 79.57\% of note-level F1-score (pitch and onset) on the Configuration-II test set of the MAPS dataset \cite{kelz2016potential}.   

\subsubsection{Multi-instrument polyphonic transcription} 
The multi-instrument transcription model is similar to the piano solo model, but its output supports 11 classes of instruments, namely piano, violin, viola, cello, flute, horn, bassoon, clarinet, harpsichord, contrabass, and oboe, which are the 11 instrument classes provided by the training dataset, MusicNet \cite{thickstun2017learning}. This model by default supports the challenging \emph{instrument-agnostic transcription} scenario, which means that the instrument classes existing in the test music piece are unknown \cite{wu2020multi}. The model achieves multi-instrument transcription by outputting 11 channels of piano rolls, each of which represents one class of instrument. The time and pitch resolution of this model is the same as the piano solo transcription model. To the best of our knowledge, this model is the first one to provide a solution of note streaming (NS) and multi-pitch streaming (MPS). When being evaluated on the MusicNet test set, the model is firstly reported to achieve a note-level F1-score of 66.59\% for the note streaming task. Both piano solo and multi-instrument transcription can be performed with the command
\texttt{omnizart music transcribe [OPTIONS] INPUT\_AUDIO}, in which user can select the pre-trained models by \texttt{[OPTIONS]}; see the documentation for details.

\subsubsection{Drum transcription} The drum transcription tool in Omnizart is a re-implementation of \cite{wei2021improving}. The model is a convolutional neural network (CNN)-based model designed to predict the onsets of percussive events from a given input audio. It comprises five convolutional layers, one attention layer, and three fully-connected layers, and the total number of parameters is around 9.4M. 
Since the onsets of percussive events are highly correlated with beats, we use an automatic beat-tracker in the data pre-processing pipeline to process the input spectrogram. The processed input, which contains rich beat information, is then fed into the model for onset prediction. To train the model, we use the self-created Audio-to-MIDI Drum dataset (A2MD), which contains 1,454 polyphonic music tracks downloaded from YouTube, and the total length is around 34 hours \cite{wei2021improving}. The time resolution of the output is 10ms. 
Evaluation on two commonly used benchmark datasets (ENST \cite{gillet2006enst} and MDB-Drums \cite{southall2017mdb} datasets) shows that the proposed model achieves a note-level F1-score of 74\% on ENST and 71\% on MDB-Drums, both of which outperform previous state-of-the-art methods.
Drum transcription in Omnizart can be performed with the command \texttt{omnizart drum transcribe [OPTIONS] INPUT\_AUDIO}.

\subsubsection{Vocal transcription in polyphonic music} The vocal transcription model is a hybrid network composed of a frame-level pitch extraction model and a note segmentation model, which inputs a multi-channel feature consisting spectrum, generalized cepstrum, and generalized cepstrum of spectrum (GCoS) \cite{wu2018automatic} extracted from the given audio, and outputs the transcribed MIDI result. For pitch extraction, we used a pre-trained Patch-CNN network proposed in \cite{su2018vocal}. For note segmentation, we improved the model proposed in \cite{fu2019hierarchical}, which is a fully supervised trained ResNet-18, by switching the model into PyramidNet-110 with ShakeDrop regularization \cite{yamada2019shakedrop} and further trained with Virtual Adversarial Training (VAT) \cite{miyato2018virtual} for semi-supervised learning. The pitch extraction model has 175K parameters while the note segmentation model has 28.49M parameters. Our model used TONAS \cite{mora2010characterization} as labeled data and MIR1K \cite{hsu2009improvement} as unlabeled data, and achieved a note transcription F1-score of 68.4\% on ISMIR2014 \cite{molina2014evaluation} dataset and outperforms all the previous state-of-the-art methods. Vocal transcription in Omnizart can be performed with the command \texttt{omnizart vocal transcribe [OPTIONS] INPUT\_AUDIO}.

\subsubsection{Chord recognition} The harmony recognition function of Omnizart is implemented using the Harmony Transformer (HT), which is a deep learning model for harmony analysis \cite{chen2019harmony}. Based on an encoder-decoder architecture, the HT model jointly recognizes the chord changes and the chord progression of an input music. Specifically, the encoder performs chord segmentation on the input, and subsequently the decoder recognizes the chord progression based on the segmentation result. With this novel approach, the HT demonstrated its promising capability of harmony recognition. In an experiment with evaluations on the McGill Billboard dataset \cite{burgoyne2011anexpert}, the HT outperformed the previous state of the arts \cite{chen2019harmony}. 
The original HT model supports either audio or symbolic input. The input is processed into the required format before being fed into the HT. Given an audio data, the harmony recognition function will represent the input as a non-negative-least-squares (NNLS) chromagram using the Chordino VAMP plugin \cite{mauch2010approximate}, yielding a 24-by-$T$-dimensional chromagram ($T$ denotes the total number of frames). For symbolic music data such as .MIDI or .MusicXML, the piano-roll representation will be applied. The output of the model includes 25 chord types (e.g., the 12 major chords, 12 minor chords, and ‘no chords’) for each 230ms. Chord recognition in Omnizart can be performed with the command \texttt{omnizart chord transcribe [OPTIONS] INPUT\_AUDIO}. Currently, Omnizart does not support symbolic music input as HT does. This utility will be extended in the future.

\subsubsection{Beat/downbeat tracking} The beat and downbeat tracking model in Omnizart is a reproduction of \cite{chuang2020beat}. The adopted model is different from most of the currently available open-source beat/downbeat tracking packages such as \texttt{madmom} \cite{bock2016madmom} and \texttt{librosa} \cite{mcfee2015librosa} which only support audio signal input. Instead, the adopted model is a beat and downbeat tracker for symbolic music data; this is also the first  deep-learning-based beat tracking solution which support such input. The model requires input in MIDI format, and outputs are beat and downbeat positions in seconds with a time resolution of 10ms. 
The model is mainly based on a two-layer bidirectional LSTM (BLSTM) recurrent neural network (RNN) with an optional attention mechanism and a fully-connected layer. The input features contain piano roll, spectral flux, and inter-onset interval extracted from the MIDI. The dimension of each hidden unit in the BLSTM network is 25 by default. The model follows the multi-tasking learning (MTL) framework to predict the probability values of beat and downbeat for each time step. Experiments 
on the MusicNet dataset \cite{thickstun2018invariances} with the synchronized beat annotation shows that the proposed model outperforms the state-of-the-art beat trackers which operate on synthesized audio \cite{chuang2020beat}.  
Beat/downbeat in Omnizart can be performed with the command \texttt{omnizart beat transcribe [OPTIONS] INPUT\_AUDIO}.

\section{Conclusion}
Omnizart represents the first systematic solution for the polyphonic AMT of general music contents ranging from pitched instruments, percussion instrument, to voice. In addition to note transcription, Omnizart also includes high-level MIR tasks such as chord recognition and beat/downbeat tracking.
As an ongoing project, the research group will keep refining the package and also extending the scope of transcription in the future.

\bibliographystyle{IEEEtran}
\bibliography{ref}
\end{document}